# Capacity of The Discrete-Time Non-Coherent Memoryless Gaussian Channels at Low SNR

Z. Rezki and David Haccoun
École Polytechnique de Montréal,
Email: {zouheir.rezki,david.haccoun}@polymtl.ca

François Gagnon
École de technologie supérieure,
Email: francois.gagnon@etsmtl.ca

***Abstract*—We address the capacity of a discrete-time memoryless Gaussian channel, where the channel state information (CSI) is neither available at the transmitter nor at the receiver. The optimal capacity-achieving input distribution at low signal-to-noise ratio (SNR) is precisely characterized, and the exact capacity of a non-coherent channel is derived. The derived relations allow to better understanding the capacity of non-coherent channels at low SNR. Then, we compute the non-coherence penalty and give a more precise characterization of the sub-linear term in SNR. Finally, in order to get more insight on how the optimal input varies with SNR, upper and lower bounds on the non-zero mass point location of the capacity-achieving input are given.**

## I. INTRODUCTION

In wireless communication, the channel estimation at the receiver is not often possible due, for instance, to the high mobility of the sender or the receiver or both. Therefore achieving reliable communication over fading channels where the channel state information (CSI) is available neither at the transmitter nor at the receiver, is of a particular interest. When CSI is not available at both ends, computing the channel capacity, known as the non-coherent capacity, as well as computing the optimal input distribution achieving this capacity, for both SISO and MIMO channels, was addressed by Marzetta and Hochwald using a block fading channel [1]. The non-coherent capacity was also computed as a function of the number of transmit and receive antennas as well as the coherence time at high SNR in [2]. At a low SNR regime, it was also shown that to a first order of magnitude of the SNR, there is no capacity penalty for not knowing the channel at the receiver which is not the case at the high SNR regime [2] [3], [4]. Recently, this power efficiency at a low SNR regime or equivalently at a large channel bandwidth has motivated work towards a better understanding of the non-coherent capacity at a low SNR regime [5], [6], [7] for both SISO and MIMO channels using several fading models.

For a discrete-time channel, computing the non-coherent capacity is a rather tedious task [8] [9]. The main difficulty in computing the non-coherent capacity relies on the fact that the capacity-achieving input distribution is discrete with a finite number of mass points, where one of them is located at the origin. The number of these mass points increases with the signal-to-noise ratio (SNR). Since no bound on the number of mass points with respect to SNR is actually available, it is very difficult to find closed form expressions for both the achievable capacity and the optimal input distribution for all SNR values. Fortunately, numerical computation of the capacity and the optimal input distribution has been made possible using the Khun-Tucker condition which is a necessary and sufficient condition for optimality, for of a SISO channel [8] and for a MIMO channel [9]. In [10], the channel mutual information at low SNR is computed and thus the capacity of a discrete time non-coherent memoryless channel is obtained through numerical optimization. However, the framework presented in [10] does not provide any insight on the capacity behavior at low SNR.

In this paper, we analyze the capacity of a discrete time non-coherent memoryless Rayleigh fading SISO channel at low SNR. The main contributions of this paper are:

1) Derivation of an analytical closed form of the channel mutual information at low SNR, which may also be considered as a lower bound on the channel mutual information for an arbitrary SNR value.
2) Derivation of a fundamental relation between the capacity-achieving input distribution and the SNR value, from which an exact capacity expression is deduced at low SNR.
3) Derivation of upper and lower bounds on the non-zero mass point location of the optimal input, which allow to deduce lower and upper bounds respectively on the non-coherent capacity at low SNR.

The paper is organized as follows. Section II presents the system model. In section III, we derive a closed form expression of the channel mutual information at low SNR which is also a lower bound on the channel mutual information at all SNR values. The optimal input distribution as well as the non-coherent capacity are computed in Section IV. Numerical results are reported in Section V and Section VI concludes the paper.

## II. CHANNEL MODEL

We consider a discrete-time memoryless Rayleigh-fading channel given by:

$$r(l) = h(l)s(l) + w(l), \qquad l = 1, 2, 3, ... \tag{1}$$

where $l$ is the discrete-time index, $s(l)$ is the channel input, $r(l)$ is the channel output, $h(l)$ is the fading coefficient and $w(l)$ is an additive noise. More specifically, $h(l)$ and $w(l)$ are independent complex circular Gaussian random variables with mean zero and variances $\sigma_h^2$ and $\sigma_w^2$, respectively. The input $s(l)$ is subject to an average power constraint, that is $E[|s(l)|^2] \leq P$, where $E[.]$ indicates the expected value. It is assumed that the channel state information is available neither at the transmitter nor at the receiver. However, even though the exact values of $h(l)$ and $w(l)$ are not known, their statistics are, at both ends. Since the channel defined in (1) is stationary and memoryless, the capacity achieving statistics of the input $s(l)$ are also memoryless, independent and identically distributed (i.i.d). Therefore, for simplicity we may drop the time index $l$ in (1). Consequently, the distribution of the channel output $r$ conditioned on the input $s$ can be obtained after averaging out the random fading coefficient $h$, yielding:

$$f_{r|s}(r|s) = \frac{1}{\pi(\sigma_h^2|s|^2 + \sigma_w^2)} \exp\left[\frac{-|r|^2}{\sigma_h^2|s|^2 + \sigma_w^2}\right]. \tag{2}$$

Noting that in (2), the conditional output distribution depends only on the squared magnitudes $|s|^2$ and $|r|^2$, we will no longer be concerned with complex quantities but only with their squared magnitudes. Let

$$I_{LB}(x_1,a)=\begin{cases} a-a\left[\frac{\ln(1+x_1^2)}{x_1^2}+\frac{1}{1+x_1^2}+\frac{x_1^2}{1+x_1^2}\cdot {}_2F_1\left(1,\frac{1}{x_1^2},1+\frac{1}{x_1^2},-\frac{(1+x_1^2)(x_1^2-a)}{a}\right)\right]-\ln\left(1-\frac{a}{x_1^2}\right)-\ln\left(1+\frac{a}{(1+x_1^2)(x_1^2-a)}\right) & \text{if } x_1>\sqrt{a}, \\ 0 & \text{if } x_1=\sqrt{a} \end{cases} \tag{9}$$

$$x_1^2-(1+x_1^2)\ln(1+x_1^2)-\pi\left(\frac{a}{x_1^2+x_1^4}\right)^{\frac{1}{x_1^2}}\csc\left(\frac{\pi}{x_1^2}\right)\left[1+x_1^2-\pi\cot\left(\frac{\pi}{x_1^2}\right)+\ln\left(\frac{a}{x_1^2+x_1^4}\right)\right]=0. \tag{11}$$

---

$y=|r|^2/\sigma_w^2$ and let $x=|s|\sigma_h/\sigma_w$. Conditioned on the input, $y$ is chi-square distributed with two degrees of freedom:

$$f_{y|x}(y|x)=\frac{1}{(1+x^2)}\exp\left[\frac{-y}{1+x^2}\right], \tag{3}$$

with the average power constraint $E[x^2]\le a$, where $a=P\sigma_h^2/\sigma_w^2$ is the SNR per symbol time.

## III. THE CHANNEL MUTUAL INFORMATION

For the channel (3), the mutual information is given by [11]:

$$I(x;y)=\int\int f_{y|x}(y|x)f_x(x)\ln\frac{f_{y|x}(y|x)}{f_{(y;x)}(y;x)}dxdy. \tag{4}$$

The capacity of channel (3) is the supremum

$$C=\sup_{E[x^2]\le a} I(x;y) \tag{5}$$

over all input distributions that meet the constraint power. The existence and uniqueness of such an input distribution was established in [8]. More specifically, the optimal input distribution for channel (3) is discrete with a finite number of mass points, where one of them is necessarily null. That is, the capacity (5) is expressed by

$$C=\max_{E[x^2]\le a}\sum_{i=0}^{N-1}p_i\int_0^\infty f_{y|x_i}(y|x_i)\ln\left[\frac{f_{y|x_i}(y|x_i)}{\sum_j p_j f_{y|x_j}(y|x_j)}\right]dy, \tag{6}$$

where $x_0=0<x_1<x_2\ldots<x_{N-1}$ are the mass point locations and where $p_0,p_1\ldots,p_{N-1}$ their probabilities respectively. Since we focus on the low SNR regime, we may use in (6) a discrete input distribution with two mass points, where one of them is null, to obtain the optimal capacity at low SNR [8]. Furthermore, this on-off signaling also provides a lower bound on the non-coherent capacity for all SNR values. That is, a lower bound on the capacity may be expressed by:

$$C_{LB}=\max_{E[x^2]\le a} I_{LB}(x;y), \tag{7}$$

where $I_{LB}(x;y)$ is a lower bound on the channel mutual information $I(x;y)$ given by:

$$\begin{aligned} I_{LB}(x;y)&=I_{LB}(x_1,p_1)\\ &=\sum_{i=0}^{1}p_i\int_0^\infty f_{y|x_i}(y|x_i)\ln\left[\frac{f_{y|x_i}(y|x_i)}{\sum_j p_j f_{y|x_j}(y|x_j)}\right]dy, \end{aligned}$$

and the average constraint power becomes: $p_1x_1^2\le a$. Note that the optimization problem in (7) deals with only two unknowns $p_1$ and $x_1$. Furthermore, it is proven below that further simplifications can be obtained, using the fact that $I_{LB}(x_1,p_1)$ is monotonically increasing in $x_1$ and thus the problem at hand may be reduced to a simpler maximization problem without constraint. We summarize this result in Lemma 1.

*Lemma 1:* The optimal capacity at low SNR and a lower bound on it for all SNR values is given by:

$$C_{LB}=\max_{x_1\ge\sqrt{a}} I_{LB}(x_1,a), \tag{8}$$

where $I_{LB}(x_1,a)$ is the channel mutual information for a given mass point location $x_1$ and a given SNR value $a$, given by (9) at the top of the page, and where ${}_2F_1(\cdot,\cdot,\cdot,\cdot)$ is the Gauss hypergeometric function. Furthermore, the constraint power holds with equality and we have: $E[x^2]=p_1x_1^2=a$.

*Proof:* The proof is presented in [12]. ■

In Lemma 1, the existence of a maximum for a given SNR value $a$ is guaranteed by the continuity of $I_{LB}(x_1,a)$ and the fact that it is bounded with respect to $x_1$ over the interval $[\sqrt{a},\infty[$. The existence of such a maximum was rigorously established in [12]. Clearly, the maximization (8) is reduced to solving the equation $\frac{\partial}{\partial x_1}I_{LB}(x_1,a)$ for a given SNR value $a$. Ideally, an analytical solution would provide an insight as to how the non-coherent capacity and the on-off signaling vary with the SNR. However, solving such an equation for arbitrary SNR values is very ambitious since it comes to finding an analytical solution to involved transcendental equations. Nevertheless, it is of interest to focus on the low SNR regime to get the benefit of some advantageous simplifications in order to elucidate the non-coherent capacity behavior at low SNR.

## IV. NON-COHERENT CAPACITY AT LOW SNR

In this section, we will use Lemma 1 to derive a fundamental analytical relation between the optimal input distribution at a low SNR regime and the particular SNR value $a$. We show in Theorem 1 that this fundamental relation holds up to an order of $a$ strictly less than 2. As is shown below, the derived relation is very useful since it allows computing the optimal input distribution for a given SNR value $a$ while providing a rigorous characterization as to how the non zero mass point locations and their probabilities vary with $a$. Moreover, the derived relation may be used to compute the exact non-coherent capacity at low SNR values.

### *A. A fundamental relation between the optimal input and the SNR*

We present the fundamental relation between the optimal input distribution and the SNR value in the following theorem.

*Theorem 1:* At a low SNR value $a$, the optimal input probability distribution for an order of magnitude of $a$ strictly less than 2 ($o(a^{2-\epsilon})$ for an arbitrary positive $\epsilon$), is given by:

$$f_x(x)=\begin{cases} x_1 & \text{with probability } p_1=\frac{a}{x_1^2}, \\ 0 & \text{with probability } p_0=1-p_1, \end{cases} \tag{10}$$

where $x_1$ is the solution of the equation (11) given at the top of the page. Furthermore, the non-coherent channel capacity is given by:

$$C(a,x_1)=a-a\cdot\frac{\ln(1+x_1^2)}{x_1^2}-a^{1+\frac{1}{x_1^2}}\cdot\frac{\pi\csc\left(\frac{\pi}{x_1^2}\right)\left(\frac{1}{x_1^2+x_1^4}\right)^{\frac{1}{x_1^2}}}{1+x_1^2} \tag{12}$$

*Proof:* The proof is presented in [12]. ■

Clearly, (11) is also a transcendental equation, for which determining an analytical solution is a very tedious task. Although it is very involved to derive an analytical solution of (11) in the form of $x_1=f(a)$, it is of interest from an engineering point of view, to

$$a = \exp\left[x_1^2 W\big(k, \varphi(x_1)\big) - x_1^2 + \pi \cot\big(\frac{\pi}{x_1^2}\big) + \ln(x_1^2) + \ln(1+x_1^2) - 1\right], \tag{14}$$

$$\varphi(x) = -\frac{\sin(\frac{\pi}{x^2})(-x^2 + \ln(1+x^2) + x^2\ln(1+x^2))}{\pi x^2} \cdot \exp\left(\frac{-\pi\cot\big(\frac{\pi}{x^2}\big)}{x^2} + 1 + \frac{1}{x^2}\right). \tag{15}$$

resolve (11) numerically and obtain the optimal $x_1$ for a given SNR value $a$. One may then get the value of the non-coherent capacity from (12). Moreover, (11) provides some insight on the behavior of $x_1$ as $a$ tends toward zero. For example, using (11), one may determine the limit of $x_1$ as $a$ tends toward zero. To see this, let $M$ be this limit and let us assume that $M$ is finite. From [6], [8], we know that for the optimal input distribution, the non-zero mass point location $x_1$ is greater than one. Thus, its limit as $a$ tends toward zero is greater or equal than one: $M \geq 1$. Then, taking the limits on both sides of (11) as $a$ goes to zero yields:

$$M^2 - (1+M^2)\ln(1+M^2) = 0. \tag{13}$$

That is, if $M$ is finite, it would be equal to zero, the unique solution to (13), but this is impossible since $M \geq 1$. Hence, consistently with [6], [8], $\lim_{a\to 0} x_1 = \infty$.

Furthermore, we have found that (11) may be written more conveniently as (14) at the top of the page, with $k = -1$ if $a \leq a_0$ and $k = 0$ elsewhere, and where $W(\cdot,\cdot)$ is the Lambert function, with $\varphi(x)$ given by (15) at the top of the page too. Also, $a_0$ is the solution of (11) for $x_1 = x_0$, where $x_0$ is the root of the equation $\varphi(x) = -\frac{1}{e}$. The number $-\frac{1}{e}$ comes out in our analysis from the fact that it is the unique point shared by the principal branch of the Lambert function $W(0,x)$ and the branch with $k=-1$, $W(-1,x)$. That is $W(0,-\frac{1}{e}) = W(-1,-\frac{1}{e})$. This guarantees the continuity of $a$ in (14) for all $x_1$ values. Numerically, we have found that $a_0 = 0.0582$ and $x_0 = \sqrt{3.93388}$. Hence, (14) may also be viewed as a fundamental relation between the optimal input distribution and the SNR $a$ for discrete-time non-coherent memoryless Rayleigh fading channels at low SNR. On the other hand, (14) provides the global answer as to how the non-zero mass point location of the optimal on-off signaling and the SNR are linked together. For this purpose, a simple analysis of (14) has been done and some important results are recapitulated in the following corollary.

*Corollary 1:* According to (14), we have:

1) For all $a \leq a_0$, $a_0 = 0.0582$, $a$ is an decreasing function with respect to $x_1$ and for all $a > a_0$, $a$ is an increasing function of $x_1$.
2) For all $a$, $x_1 \geq x_0$, where $x_0 = \sqrt{3.93388}$.
3) $\lim_{x_1\to\infty} a = 0$.

Corollary 1 agrees with [8] where it was shown using computer simulation that the non-zero mass point location passes through a minimum before moving upward. However, by specifying the edge point $(x_0, a_0)$, Corollary 1 gives a more precise characterization concerning this peculiar behavior of the non-zero mass point locations. Furthermore, Corollary 1 also refines the lower bound on $x_1$, $x_1 > 1$ and derives $x_0$ as an improved lower bound on the non-zero mass point location at low SNR. Moreover, from (14), we may write:

$$\ln(a) + x_1^2 = x_1^2 W\big(k, \varphi(x_1)\big) + \pi\cot\big(\frac{\pi}{x_1^2}\big) + \ln\big(x_1^2(1+x_1^2)\big) - 1. \tag{16}$$

It is then easy to check that the right hand side (RHS) of (16) is a decreasing function of $x_1$ for $x_1 < x_0$, which yields an upper bound on $x_1$:

$$x_1^2 \leq -\ln(a) + \xi_0, \tag{17}$$

where $\xi_0 = \ln(a_0) + x_0^2$, which is again consistent with the upper bound derived in [6]. Note that the upper bound (17) is valid for all $a \leq a_0$ whereas the upper bound provided in [6] holds for $a \ll a_0$ for which $\xi_0$ is negligible. On the other hand, combining (17) and the lower bound on $x_1$ provided in Corollary 1 one may obtain:

$$a^\alpha x_0^2 \leq a^\alpha x_1^2 \leq a^\alpha(\xi_0 - \ln(a)). \tag{18}$$

for all $\alpha > 0$. That is:

$$\lim_{a\to 0}\big(a^\alpha x_1^2\big) = 0, \tag{19}$$

which means that $a^\alpha$ tends toward zero faster than $x_1^2$ does toward infinity. This result may also be used to gain further insight on the capacity behavior at low SNR. For instance, from (12), we may write the non-coherent capacity as:

$$C(a) = a + o(a), \tag{20}$$

where $o(a) = -a \cdot \frac{\ln(1+x_1^2)}{x_1^2} - a^{1+\frac{1}{x_1^2}} \cdot \frac{\pi\csc\left(\frac{\pi}{x_1^2}\right)\left(\frac{1}{x_1^2+x_1^4}\right)^{\frac{1}{x_1^2}}}{1+x_1^2}$, meaning that the non-coherent capacity varies linearly with $a$ at low SNR and hence non-coherent communication at low SNR may be qualified as energy efficient communication.

### *B. Energy efficiency and non-coherence penalty*

In general, the capacity of a channel including a Gaussian channel and a Rayleigh channel varies linearly at low SNR [6]. The difference between these channels in terms of capacity can only be explained by the sub-linear term $o(a)$ in (20). The sub-linear term has been defined in [6] as:

$$\Delta(a) := a - C(a). \tag{21}$$

At low SNR, the sub-linear term $\Delta(a)$ is also related to the energy-efficiency. Let $E_n$ be the transmitted energy in Joules per information nat, then we have:

$$\frac{E_n}{\sigma_w^2} \cdot C(a) = a. \tag{22}$$

Using (21), we can write:

$$\frac{E_n}{\sigma_w^2} = \frac{1}{1 - \frac{\Delta(a)}{a}} \approx 1 + \frac{\Delta(a)}{a}, \tag{23}$$

where the approximation holds if $\frac{\Delta(a)}{a}$ is sufficiently small. Note that if $\frac{\Delta(a)}{a} \to 0$, then from (21) and (23), we have respectively: $C(a) \approx a$ and $\frac{E_n}{\sigma_w^2} \approx 1$, which implies that the highest energy efficiency of -1.59 (dB) per information bit could be theoretically achieved. For a Gaussian channel and a fading channel under the coherent assumption, the sub-linear terms are respectively given by [6]:

$$\Delta_{AWGN}(a) = \frac{1}{2}a^2 + o(a^2) \tag{24}$$

$$\Delta_{coherent}(a) = \frac{1}{2}E[\|h\|^4]a^2 + o(a^2) \tag{25}$$

For a non-coherent Rayleigh fading channel, the sub-linear term can be computed using (12):

$$\Delta(a) = a \cdot \frac{\ln(1+x_1^2)}{x_1^2} + a^{1+\frac{1}{x_1^2}} \cdot \frac{\pi\csc\left(\frac{\pi}{x_1^2}\right)\left(\frac{1}{x_1^2+x_1^4}\right)^{\frac{1}{x_1^2}}}{1+x_1^2}. \tag{26}$$

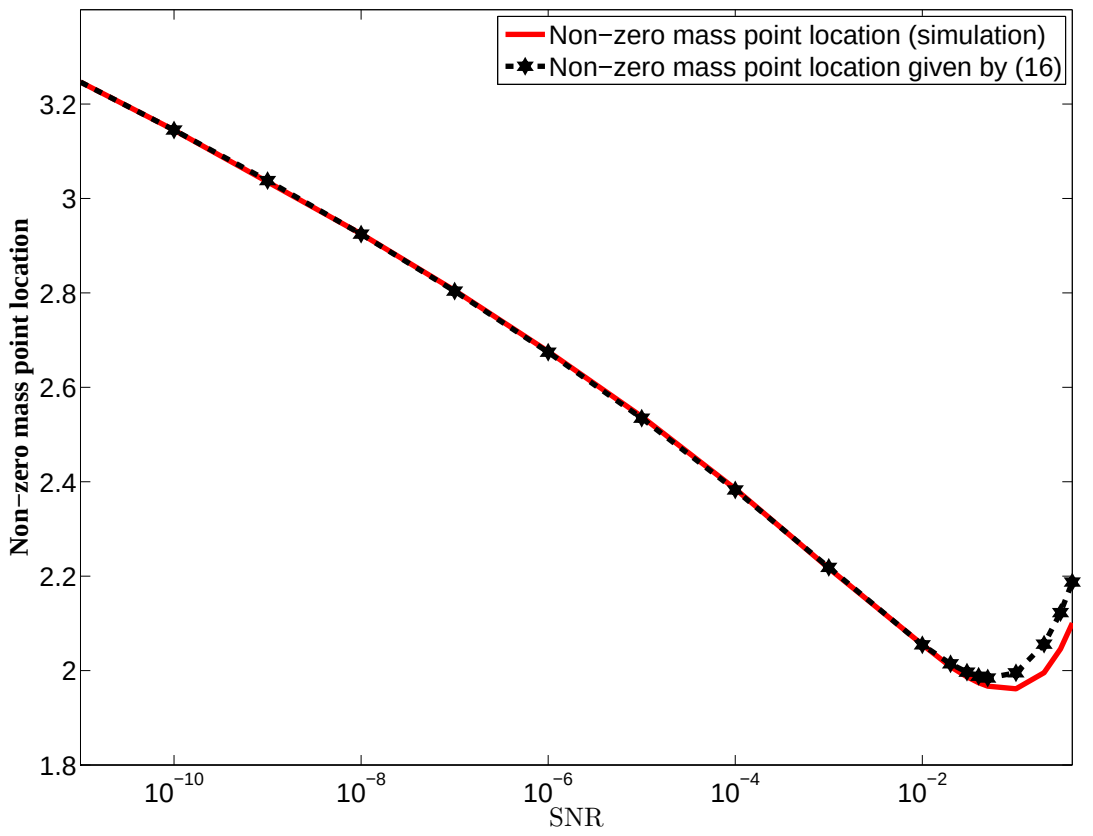

Figure 1. Location of non-zero mass point versus the SNR value $a$ (linear).

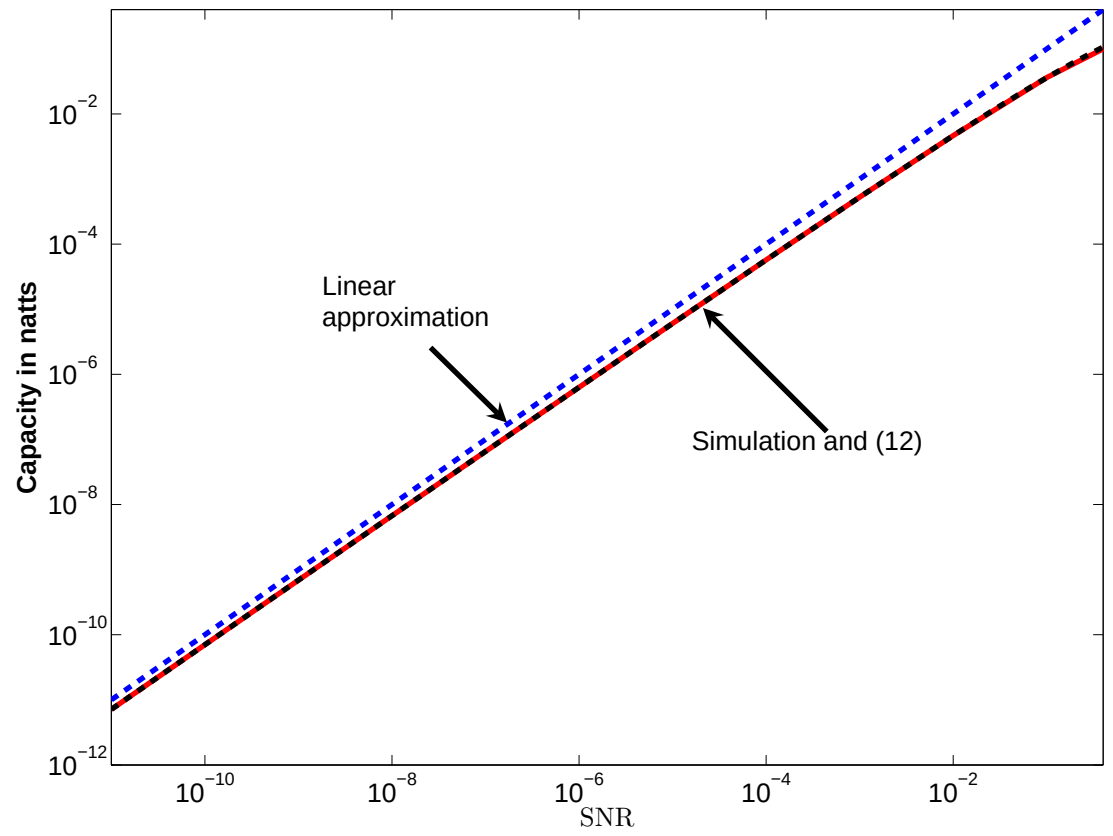

Figure 2. Non-coherent capacity versus the SNR value $a$ (linear).

Note that at very low SNR and following (26), $\frac{\Delta(a)}{a}$ converges to zero making the non-coherent Rayleigh channel also energy efficient. However, as SNR increases, the convergence of $\frac{\Delta(a)}{a}$ to zero is slower than $\frac{\Delta_{AWGN}(a)}{a}$ and $\frac{\Delta_{coherent}(a)}{a}$. This could be seen from (19) indicating that $x_1$ converges slower to infinity than $a$ does to zero.

In the range of SNR values of interest, we may define the non-coherence penalty per SNR as:

$$\frac{C_{coherent}(a) - C(a)}{a}. \tag{27}$$

where $C_{coherent}$ is the channel capacity under coherent assumption. Now, from [6], we can write $C_{coherent}$ as:

$$C_{coherent}(a) = a + O(a) = a + o(a^{2-\alpha}), \tag{28}$$

for any $1 > \alpha > 0$. Recalling that the non-coherent capacity in (12) was obtained using series decomposition to an order strictly smaller than 2 ($o(a^{2-\epsilon})$ for an arbitrary positive $\epsilon$), then combining (12) and (28), we derive the exact non-coherence penalty per SNR up to this order:

$$\begin{aligned}\frac{C_{coherent}(a) - C(a)}{a} &= \frac{C_{coherent} - C}{C_{coherent}} \\ &= \frac{\ln(1+x_1^2)}{x_1^2} + a^{\frac{1}{x_1^2}} \frac{\pi \csc\left(\frac{\pi}{x_1^2}\right)\left(\frac{1}{x_1^2+x_1^4}\right)^{\frac{1}{x_1^2}}}{1+x_1^2}\end{aligned} \tag{29}$$

Note that the non-coherence penalty is equal to the sub-linear term $\Delta(a)$. Now using (19), dividing both sides of (29) by $a^\alpha$, ($\alpha > 0$) and taking the limit as $a$ tends to zero yields:

$$C_{coherent}(a) - C(a) \gg a^{1+\alpha}, \tag{30}$$

where $\gg$ means: $\lim_{a\to 0} \frac{C_{coherent}(a)-C(a)}{a^{1+\alpha}} = \infty$. Inequality (30) indicates that not only non-coherence penalty is much greater than $a^2$ as was established in [5], but more precisely, it is much greater than $a^{1+\alpha}$ since $a^{1+\alpha} \gg a^2$, $1 > \alpha > 0$. Again, this result is in full agreement with [6].

In this subsection, we have discussed exact closed forms of the optimal input distribution and the non-coherent capacity based on the fundamental relation (11) or equivalently (14). However, one may be interested in deriving simpler lower and upper bounds on these quantities in order to provide more insight on how they vary with the SNR value $a$. This is discussed next.

### C. *Upper and lower bounds on the non-coherent capacity*

Considering (14), since we are interested in the low SNR regime, we assume for simplicity that $a \le a_0$. Thus the Lambert function in (14) is the branch with $k = -1$, $W(-1, x)$. A lower bound on the non-coherent capacity is easily obtained by combining (17) and (12) and will be referred to as $C_{LB}(a)$. We now derive the lower bound on the optimal non-zero mass point location and the upper bound on the non-coherent capacity in Theorem 2.

*Theorem 2:* At low SNR values $a$, a lower bound on the optimal non-zero mass point location is given by:

$$x_{1,LB} = \frac{y}{\sqrt{-W\left(-1, \varphi\left(\frac{y}{-\ln\left(-\varphi(y)\right)}\right)\right)}}, \tag{31}$$

where $y = \sqrt{1 + \ln \frac{1}{a}}$. Furthermore, an upper bound on the non-coherent capacity can be obtained from (12) as:

$$C_{UB}(a) = C(a, x_{1,LB}) \tag{32}$$

*Proof:* The proof is presented in [12]. ■

## V. Numerical Results and Discussion

The curves in Fig. 1 show respectively, the non-zero mass point location of the capacity-achieving input distribution $x_1$ obtained using maximization (8), and the one obtained using relation (11) or equivalently (14). As can be seen from Fig. 1, the two curves are undistinguishable at low SNR, confirming that (15) is exact at low SNR. As the SNR increases, a small discrepancy between the two curves starts to appear. This is expected since (14) holds for up to an order of magnitude strictly smaller than 2 and thus for small SNR values, (but not smaller than about $2.10^{-2}$), a discrepancy may appear. Nevertheless, even for an SNR greater than $2.10^{-2}$, the curve obtained using (14) is very instructive especially as it follows the same shape as the one obtained by simulation results. An interesting future work would be to use (15) in order to understand why a new mass point should appear as the SNR increases. It should be mentioned that the discrepancy observed in Fig. 1 may be rendered as small as desired using high order series expansion. However, the analysis would be unrewardingly too complex.

Figure 2 depicts the non-coherent capacity curves. Again, the curve obtained by computer simulation and the one obtained using (12)

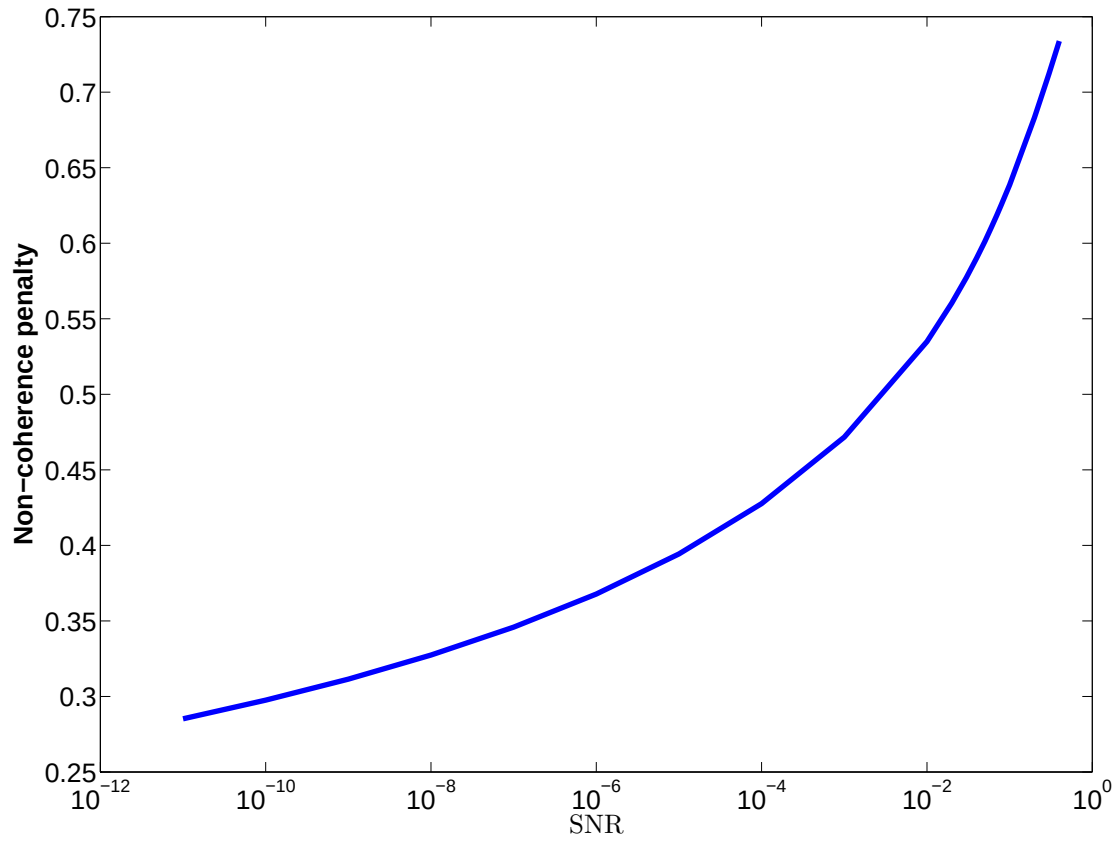

Figure 3. Non-coherence penalty per SNR versus the SNR value $a$ (linear).

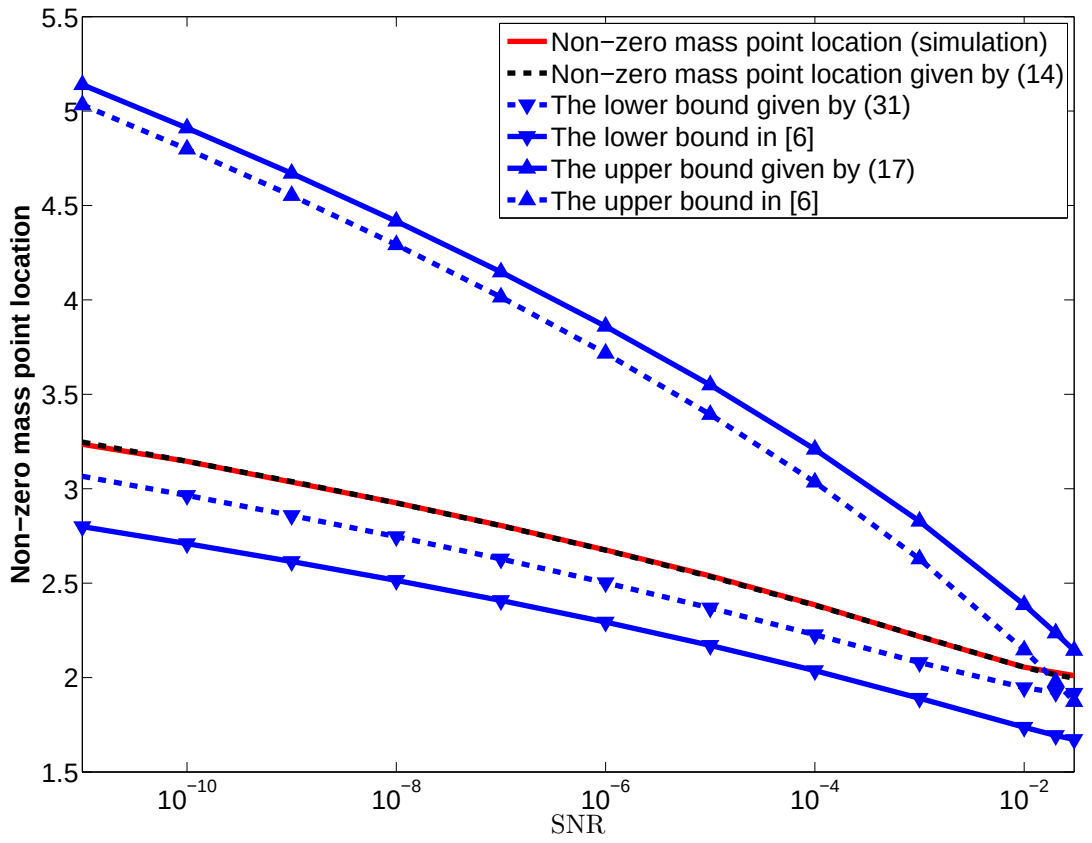

Figure 4. Exact non-zero mass point locations and the derived upper and lower bounds as well as those reported in [6] versus the SNR value $a$ (linear).

are undistinguishable. More interestingly, the discrepancy observed at not very low SNR values in Fig. 1 has vanished, implying that the capacity is not very sensitive to the non-zero mass point location. Also shown in Fig. 2 is the linear approximation $C(a) = a$, which is an upper bound on the capacity. As can be noticed in Fig. 2, the linear approximation follows the same shape as the exact non-coherent capacity curves at low SNR and becomes quite loose for SNR values greater than $10^{-2}$. This implies that the sub-linear term defined in (21) is much more important at these SNR values. This can be seen in Fig. 3 where we have plotted the non-coherence penalty percentage given by (29). Figure 3 confirms that there is no substantial gain in the channel knowledge in a capacity sense at very low SNR, thus indicating that non-coherent communication is almost as power-efficient as AWGN and coherent communications. As the SNR increases, a non-coherence penalty begins to appear reaching up to $70\%$.

The derived upper and lower bounds on the non zero mass point locations given respectively by (17) and (31) as well as the bounds derived in [6] are plotted in Fig. 4 along with the exact curves at low SNR. As can be seen in Fig. 4, the upper bound in [6], albeit tighter than (17), crosses the exact curves at about $2.10^{-2}$. At these not so low SNR values, the derived bound in [6] is no longer an upper bound, consistently with our discussion in subsection IV-A. On the other hand, the lower bound (31) is tighter than the one derived in [6] for all SNR values.

## VI. Conclusion

In this paper, we have addressed the analysis of the capacity of discrete-time non-coherent memoryless Gaussian channels at low SNR. We have computed explicitly the channel mutual information at low SNR which is also a lower bound on the channel mutual information, albeit not necessarily at low SNR values. Using the derived expression of the channel mutual information, we have been able to provide a fundamental relation between the non-zero mass point location of the capacity-achieving input distribution and the SNR. This fundamental relation brings the complete answer about how the optimal input distribution varies with the power constraint at low SNR. It also provides an analytical explanation on what was previously observed through computer simulation in [8] about the peculiar behavior of the non-zero mass point location at low SNR values. The exact non-coherent capacity has been derived and insights on the capacity behavior which can be gained through functional analysis has been shown.

In order to better understand how the non-zero mass point location varies with the SNR, we have also derived lower and upper bounds which have been compared to recently derived bounds. The newly derived lower bound is tighter for all SNR values of interest, whereas somewhat looser, the upper bound was shown to hold for larger SNR values.